\documentclass[aps,prd,twocolumn,superscriptaddress,nofootinbib]{revtex4-2}

\usepackage{graphicx}
\usepackage{dcolumn}
\usepackage{bm}
\usepackage{color}
\usepackage{hyperref}
\usepackage[english]{babel}
\usepackage{amsmath}
\usepackage{comment}
\usepackage{ulem}

\newcommand{\Msun}{M_{\odot}}
\newcommand{\ma}{m_{a}}

\newcommand{\gten}{g_{\rm{10}}}

\newcommand{\MeV}{\rm{MeV}}
\newcommand{\km}{\rm{km}}
\newcommand{\tpb}{t_{\rm{pb}}}
\newcommand{\s}{\rm{s}}

\begin{document}

\preprint{APS/123-QED}

\title{\texorpdfstring{Rotation-induced Relaxation of Supernova Constraints on Axionlike Particles}{First Line - Second Line}}


\author{Tsurugi Takata}
    \affiliation{Department of Applied Physics, Fukuoka University, 8-19-1 Nanakuma, Jonan-ku, Fukuoka 814-0180, Japan}
\author{Kanji Mori}
\affiliation{Department of Physics, Faculty of Science and Technology, Keio University, 3-14-1 Hiyoshi, Kohoku-ku, Yokohama, Kanagawa 223-8522 Japan}
\author{Ko Nakamura}
    \affiliation{College of Arts and Sciences, J. F. Oberlin University, 3758 Tokiwamachi, Machida, Tokyo 194-0213, Japan}
\author{Kei Kotake}
    \affiliation{Department of Applied Physics, Fukuoka University, 8-19-1 Nanakuma, Jonan-ku, Fukuoka 814-0180, Japan}

\begin{abstract}
We study how rotation modifies the constraints on MeV-scale axion-like particles (ALPs) coupled to photons derived from SN 1987A. We constrain the ALP parameter space based on both the energy-loss argument and the gamma-ray limits, and examine how these constraints are affected by stellar rotation. Adopting initial angular velocities of $\Omega_0=0.0$ and $1.0\,\mathrm{rad\,s^{-1}}$ in the iron core, 
we carry out two-dimensional core-collapse supernova simulations for three progenitor models --- a $14+9\,\Msun$ binary and $13\,\Msun$ and $18\,\Msun$ single stars with solar metallicity --- and estimate ALP emission rates through post-processing.
We find that rotation suppresses ALP emission by reducing the core temperature via centrifugal support. Rotation also reduces the neutrino luminosity, but the suppression of ALP emission is more effective, leading to relaxed constraints within a criterion based on the energy-loss argument.
This relaxation is particularly pronounced in the rotating $18\,\Msun$ model, where a substantial decrease in the central temperature occurs at $\tpb=0.8-1\,\s$. 
In this criterion, such rapid temporal variations in temperature indicate that the resulting constraints depend sensitively on both the evaluation time and the underlying supernova model.
For a gamma-ray limit from the SN~1987A observation, rotation has a negligible impact on the constraint. This is because the ALP-induced gamma-ray fluence observed at Earth is proportional to the fourth power of the ALP-photon coupling constant, making the constraint relatively insensitive to the rotational suppression of ALP emission.

\end{abstract}
\maketitle

\section{Introduction}
Core-collapse supernovae (CCSNe) represent one of the most powerful astrophysical events in the Universe, leaving behind a neutron star or a black hole. During the core collapse, their central cores reach a temperature of $\sim 50\,\text{MeV}$ and a density of $\sim 10^{15}\,\text{g cm}^{-3}$. In such extreme environments, feebly interacting particles beyond the Standard Model, such as axion-like particles (ALPs\footnote{Axion is a hypothetical particle that is introduced to solve the strong CP problem \cite{wilczek_problem_1978,weinberg_new_1978}. Axions can interact with the Standard Model particles such as nucleons, photons, and electrons, and hence they can be produced in a hot plasma in astrophysical objects. Motivated by the recent development in the string theory \cite{svrcek_axions_2006,cicoli_type_2012,arvanitaki_string_2010}, a more general class of axion-like particles (ALPs) has been introduced as a new particle beyond the Standard Model, where the mass and the coupling constants are treated as independent parameters.}), can be efficiently produced. 
In the case that ALPs interact only with photons, they can be produced in CCSNe via the Primakoff process and photon coalescence. ALPs can influence explosion properties from CCSN events such as the explosion energy, neutrino and gravitational-wave signals, and protoneutron star (PNS) properties, since they can contribute to energy transport in the star.
By comparing predictions with observational data, constraints can be placed on the ALPs parameter space. Hence, CCSNe have been regarded as one of useful astrophysical laboratories for probing ALPs (see Refs. \cite{raffelt_astrophysical_2008,Caputo_astropyhsical_2024,Carenza_axion_2025} for recent reviews).

In the standard CCSN scenario, almost all of the gravitational binding energy released during core collapse is carried away by neutrinos. 
If ALPs are produced in the core environment of a CCSN, they would introduce an additional channel of energy transport within the CCSN.
When the additional energy-loss mechanism competes with or exceeds neutrino cooling, it substantially shortens the duration of the neutrino burst. If the neutrino-burst duration predicted by the simulations is inconsistent with the observed signal, ALPs that lead to such excessive cooling are excluded. This argument is known as the ``energy-loss argument". Observationally, the neutrino burst from SN~1987A was detected for $\sim 12\,\s$ \cite[e.g.][]{pagliaroli_improved_2009}, which provides a stringent constraint on the parameters of ALPs \cite{ellis_constraints_1987,Turner_axion_1988,raffelt_bounds_1988,mayle_constraints_1988,mayle_updated_1989,burrows_axions_1989,keil_fresh_1997,Chang_supernova_2018,bollig_muons_2020,carenza_constraints_2020,ertas_on-the-interplay_2020,lucente_supernova_2021,caputo_muonic_2022,lucente_constraining_2022,Ferreira_strong_2022,2024PhRvD.109b3001L,Haxton_isthe_2026,Haxton_revisiting_2026,2025arXiv250905973M}.

In addition to the neutrino observations, SN~1987A gamma-ray measurements also provide severe constraints on ALPs. If ALPs are produced inside supernovae (SNe) and most of them escape the star, they travel through space and can convert into photons, producing a gamma-ray signal excess that cannot be explained by standard SN models.  In fact, the Gamma-Ray Spectrometer (GRS) on board the Solar Maximum Mission (SMM) was operational during the neutrino burst from SN~1987A. However, it did not capture any evidence of the excess \cite{chupp_experimental_1989}. This non-detection sets an upper limit on the fluence of \(25-100\) MeV photons from ALPs, thus imposing constraints on their parameters \cite{BROCKWAY_SN1987A_1996,Grifols_Gamma_1996,giannotti_new_2011,payez_revisiting_2015,jaeckel_decay_2018,Ferreira_strong_2022,caputo_muonic_2022,diamond_axion-sourced_2023,Hoof_Updated_2023,muller_investigating_2023}.

For current studies of ALP constraints based on these arguments, one-dimensional SN simulations are the most widely used. They can track the long-term evolution of the neutrino signals and PNS properties required to estimate the ALP production rate and gamma-ray fluence.  However, over the years, numerous SN studies have demonstrated that multidimensional hydrodynamic effects play a crucial role in SN dynamics, including PNS convection \cite[e.g.][]{takiwaki_comparison_2014,Vartanyan_successful_2019,Ott_progenitor_2018,lentz_three-dimensional_2015}. 
These effects can also impact the PNS structure---such as its temperature and density---as well as neutrino luminosity. Alongside multidimensional hydrodynamic effects, rotation, which is a ubiquitous feature of stars \cite{Maeder_rotating_2012}, can also influence the dynamics, neutrino emission, and PNS properties through centrifugal support, deformation, angular momentum transport, and non-axisymmetric instabilities \cite[e.g.][]{Janka_phrsics_2016,summa_rotation_2018,takiwaki_three-dimensional_2016}. 
In particular, rotation reduces both central density and temperature of the PNS. Since the  ALP emission rate is sensitive to thermal conditions, constraints derived from the ALP production rate can be influenced by multidimensional and rotational effects. ALP constraints based on CCSN simulations incorporating these effects remain insufficiently explored. Therefore, further studies are needed to improve their reliability.

In this paper, we perform two-dimensional axisymmetric CCSN simulations with rotation for a SN~1987A progenitor model to understand the impact of rotation on ALP parameter constraints. Based on the simulation data, we evaluate the ALP production rates through post-processing analysis and revisit the MeV-scale ALP parameter space constrained by the energy-loss argument and gamma-ray limits. 
We also adopt $13\Msun$ and $18\Msun$ single-star models with solar metallicity to investigate the progenitor dependence of the rotational effects on ALP emissions.

This paper is organized as follows. In Section \ref{SN models}, we describe the setup of our simulations and the properties of SN models. In Section \ref{alp production rates}, we explain the ALP model we adopt and estimate the ALP cooling rates from the SN models. 
In Sections~\ref{energy-loss argument} and \ref{gamma-ray limit}, we explain two methods to constrain the ALP parameter space from SN, apply them to our CCSN models, and show the results. In Section \ref{CandD}, we summarize  our results and draw conclusions.

\section{SN Models}
\label{SN models}
In this study, we perform two-dimensional CCSN simulations with rotation to estimate the ALP production rate through post-processing analysis. In the present simulations, the feedback effect of ALP cooling and heating is not included, whereas our previous studies \cite{mori_shock_2022, mori_multimessenger_2023, Takata_progenitor_2025} incorporated them into the simulations. This is because performing individual two-dimensional simulations for a wide range of $m_a$ and $g_{a\gamma}$ is computationally expensive. Post-processing allows us to scan the ALP parameter space and obtain constraints.

\subsection{Numerical Setup}
\label{numerical setup}
We perform two-dimensional CCSN simulations using
the 3DnSNe code \cite{takiwaki_three-dimensional_2016}, which is a multi-dimensional neutrino radiation hydrodynamics code constructed to study CCSNe.
The neutrino transport is solved by the three-flavor isotropic diffusion source approximation (IDSA) scheme \cite{liebendoerfer_isotropic_2009, takiwaki_three-dimensional_2016}.
We use the state-of-the-art neutrino opacity set
\cite{kotake_impact_2018} and the neutrino energy spectrum is discretized into
20 energy bins for $0 < \varepsilon_\nu \leq 300 \,\MeV$. 
As a solution of the Poisson equation for gravity, $\Delta \Phi = 4\pi G\rho$, the spherically symmetric gravitational potential is taken in the form of the effective general relativistic effect \cite{marek_exploring_2006}, and the multipolar components are added following Ref. \cite{Wongwathanarat_hydrodynamical_2010}.
The nuclear equation of state (EoS) is from Ref. \cite{lattimer_generalized_1991} with incompressibility of $K=220$\,MeV. A grid resolution of \(N_r \times N_{\theta} = 720 \times 128\) is used for the radial and polar directions, and the spatial range of the computational domain is within the radial coordinate $r<20000$ km.

The progenitor we employ is the $14+9\,M_\odot$ merger model (hereafter the m14 model) from Ref. \cite{Urushibata_progenitor_2018}.
This model is based on the slow-merger scenario \cite{Podsiadlowski_merger_1990,Podsiadlowski_presupernova_1992} and successfully reproduces the progenitor characteristics of SN~1987A such as its evolutionary properties. 
Previously, this model has been studied in terms of matter mixing in the outer envelope using the FLASH code \cite{Ono_corecollapse_2020} and the dynamical evolution using the 3DnSNe code \cite{Nakamura_threedimensional_2022}. 
Additionally, we emply $13\,\Msun$ and $18\,\Msun$ solar metallicity progenitor models
from Ref. \cite{woosley_nucleosynthesis_2007} to investigate the progenitor dependence of the rotational effects on the constraints. These models are red supergiants and massive enough for their cores to undergo gravitational collapse during the final stage of their evolution \cite{smartt_detection_2004, smartt_death_2009}, leading to type II supernova explosions, which are the most commonly observed subclass of core-collapse SNe. These three progenitor models have different compactness, which leads to differences in core temperature, affecting the ALP production rate. Hence, the models examined in this study are suitable for investigating the impact of progenitor structure on ALP emission and the resulting constraints.

We assume the initial rotation in cylindrical form for our simulations as adopted in previous studies \cite[e.g.][]{Eriguchi_gravitational_1985, Zwerger_dynamics_1997, Ott_gravitational_2004}. At the beginning of our simulations, the angular velocity profile is given by
\begin{eqnarray}
\Omega(X,Z)=\Omega_{0}\frac{X_{0}^2}{X^2+X_{0}^2}\frac{Z_{0}^4}{Z^4+Z_{0}^4},
\end{eqnarray} where $X$ and $Z$ denote the distances measured from the rotation axis and the equatorial plane, respectively, and $X_0=Z_0=1000\,\mathrm{km}$ represent the characteristic scale of the rotating core in each direction \cite{kotake04}. $\Omega_0$ is the rotational velocity at center and we assume $\Omega_0 = 0.0\,\mathrm{rad\,s^{-1}}$ for non-rotating models and $\Omega_0 = 1.0\,\mathrm{rad\,s^{-1}}$ for rapidly rotating models. 

In this work, we perform six CCSN simulations for the three progenitor models with and without rotation until the post-bounce time \(\tpb=1.0\,\mathrm{s}\). These models are labeled as follows. For example, the model with the $13.0\,\Msun$ progenitor and \(\Omega_0 = 1.0\,\mathrm{rad\,s^{-1}}\) is hereafter denoted as ``s13r1.0".

\subsection{Model Properties}
\label{model properties}

\begin{figure*}[t]
    \centering
    \includegraphics[keepaspectratio, width=17cm,clip]{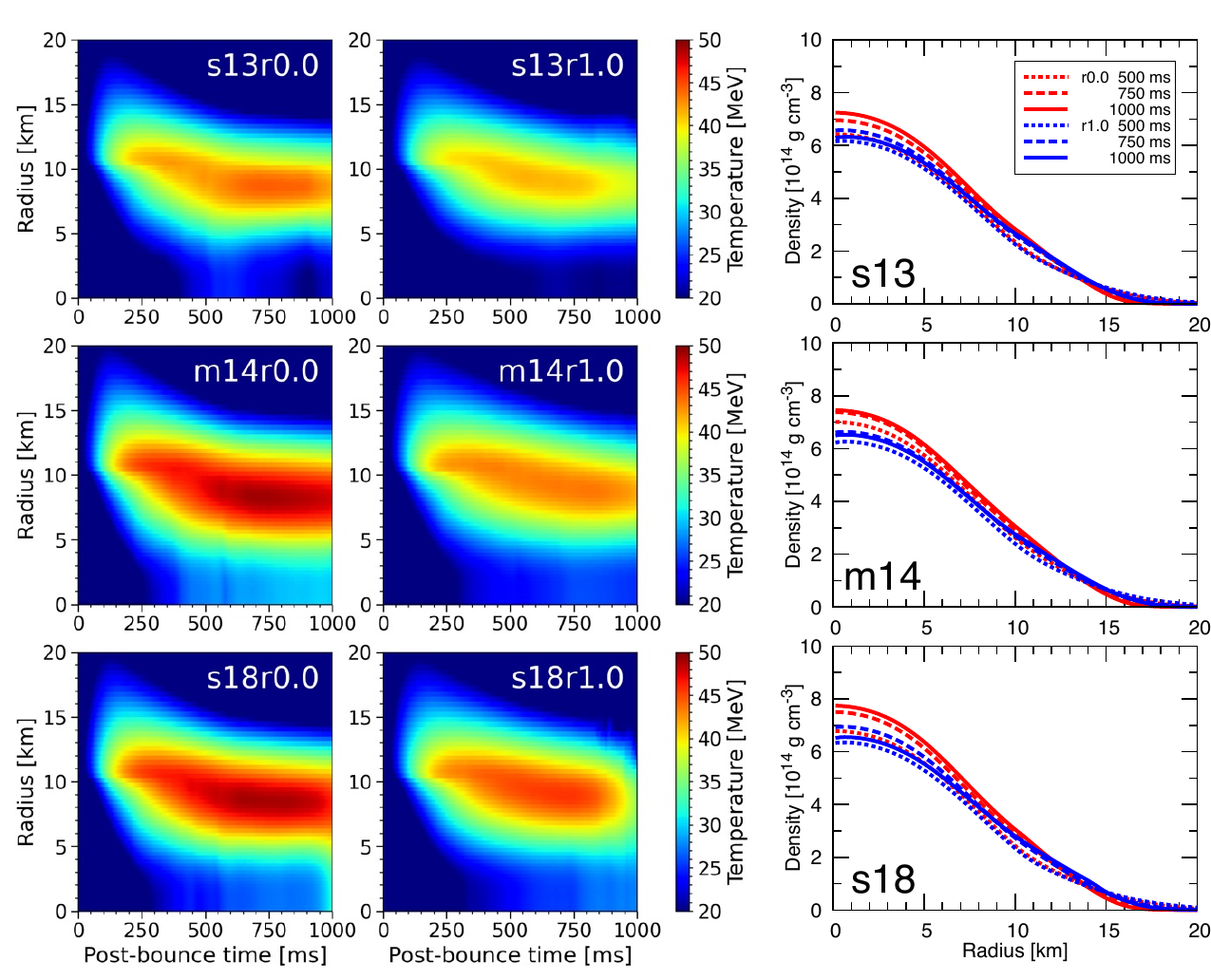}
\caption{Left and middle panels: A space-time diagram of the central temperature. In all models, the temperature peaks at\(\sim10\) km. Compared with non-rotating models, rotating models show a lower central temperature, since centrifugal forces suppress the release of gravitational binding energy. Right panels: radial profile of the central density at $500\,{\rm ms}$ (dotted lines), $750\,{\rm ms}$ (dashed lines), and $1000\,{\rm ms}$ (solid lines) after bounce. 
Rotating models (blue lines) show lower central density than non-rotating models (red lines).
\label{fig:tmp-dns}}
\end{figure*}

In the left panels of Figure \ref{fig:tmp-dns}, we show the space-time diagram of the temperature for each model, which is a critical ingredient for the ALP production rate. Within $\tpb\le1\,\s$ (with $\tpb$ the postbounce time), all models show a temperature peak at $r = 7$--$10\,\mathrm{km}$. The m14 and s18 models have relatively high compactness, resulting in higher temperatures in this region than in the s13 model. Comparing the rotating and non-rotating cases for a given progenitor model, we find that rotation leads to a lower core temperature due to the suppression of gravitational energy release by centrifugal support. In particular, at $\tpb= 1\,\s$ the s13r1.0 and s18r1.0 models exhibit a pronounced reduction in the temperature compared to the non-rotating cases.

The right panels of Figure \ref{fig:tmp-dns} show the radial density profiles in the central region at several selected times. Due to the PNS contraction over time, the central density increases and exceeds the nuclear saturation density within \(r \le 10\,\mathrm{km}\). In the rotating models (blue lines), the density is relatively lower because the centrifugal support makes the post-bounce structure less compact.

\section{ALP Production Rates}
\label{alp production rates}
\begin{figure*}[t]
    \centering
    \includegraphics[keepaspectratio, width=17cm,clip]{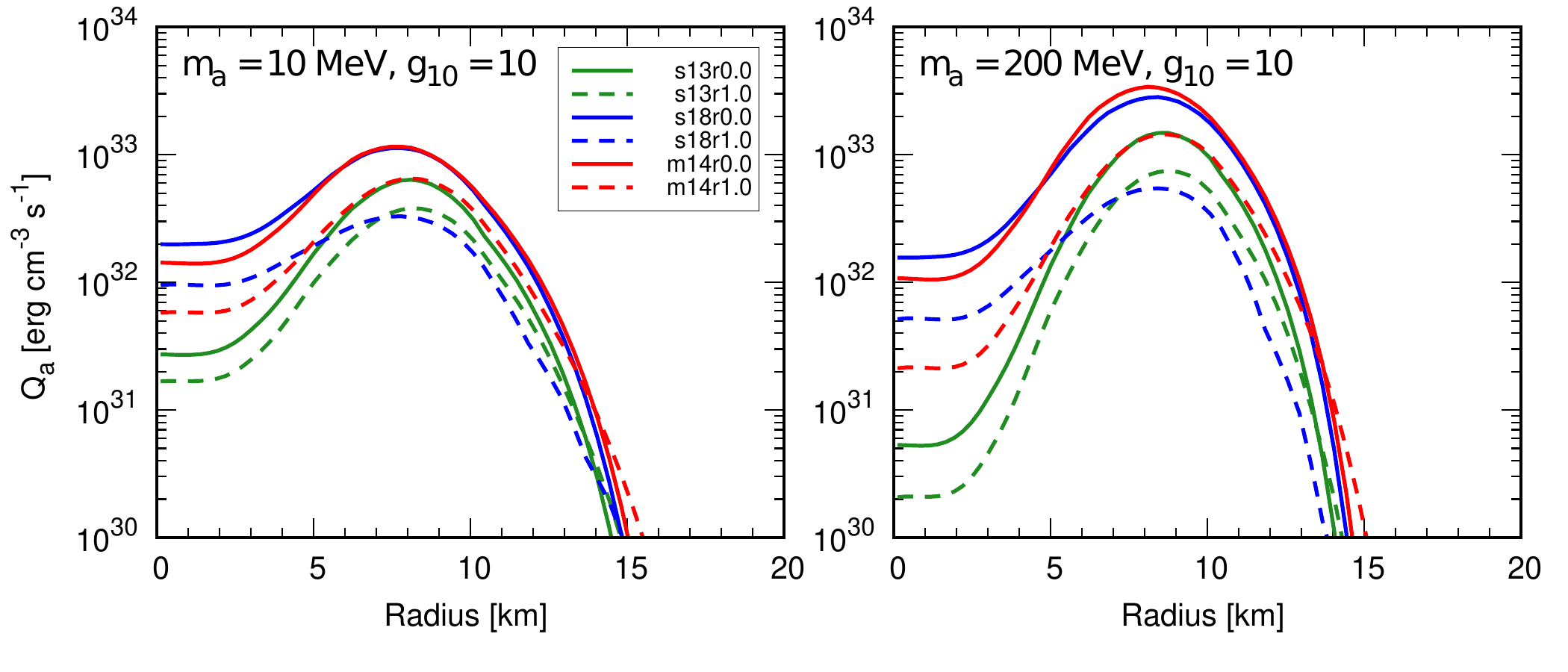}
    \caption{The radial profiles of the ALP cooling rate at $\tpb = 1$ s
    for ALP masses of 10 MeV (left panel) and 200 MeV (right panel). 
    Shown are the cases for $g_{10}=10$. In rotating models (dashed lines), the ALP production rate is lower than that in non-rotating models (solid lines). This is because rotation reduces the central core temperature, and the production rate is highly sensitive to temperature.
    }
\label{fig:Qa}
\end{figure*}

We estimate the ALP production rate based on the properties such as density, temperature, and $Y_{\rm e}$ extracted from core-collapse simulations. The ALP production rate is calculated following Ref. \cite{mori_shock_2022}, and we outline the method in this section.

We consider a photophilic ALP model where the ALPs are generated through two photon interaction processes; the Primakoff process $(\gamma + p \to a + p)$ and the photon coalescence $(\gamma + \gamma \to a)$. The Primakoff rate is given as 
\begin{equation}\label{eq:Primrate}
    \frac{d^{2} n_{a}}{dtd\omega} \Bigg|_{\rm{prim}}  = \frac{1}{\pi^2}\omega\sqrt{\omega^{2}-\omega_{\rm{pl}}^{2}}\Gamma _{\gamma \to a} f(\omega),
\end{equation} 
where $n_a$ is the number density of ALPs, $\omega$ is photon energy, $f(\omega)$ is the Bose-Einstein distribution of photons, and $\omega_{\rm{pl}}$ is plasma frequency, which is equivalent to the ``effective photon mass".
The ALP energy $E$ is equal to $\omega$ because of the energy conservation, and $\Gamma _{\gamma \to a}$ is given by \cite{di_lella_search_2000}
\begin{equation}
\begin{split}
    \Gamma _{\gamma \to a} = g^2_{a \gamma} \frac{T \kappa^2}{32 \pi} \frac{p}{E} \left(\frac{((k+p)^2+\kappa^2) ((k-p)^2+\kappa^2)}{4kp\kappa^2}\right. \times \\
    \left.\ln\left(\frac{(k+p)^2+\kappa^2}{(k-p)^2+\kappa^2}\right)- \frac{(k^2-p^2)^2}{4kp\kappa^2} \ln\left(\frac{(k+p)^2}{(k-p)^2}\right)-1\right),
\end{split}
\end{equation}
where $T$ is the temperature, $\kappa$ is the Debye-H\"{u}ckel scale, $p$ is the ALP momentum, and $k$ is the wave number of photons in plasma.

The photon coalescence rate is given as \cite{di_lella_search_2000}
\begin{equation}\label{eq:Coalrate}
    \frac{d^{2} n_{a}}{dtdE} \Bigg|_{\rm{coal}} = g^{2}_{\alpha \gamma} \frac{m^{4}_{a}}{128 {\pi}^3} p \left(1- \frac{4 {\omega}_{\rm pl}^{2}}{m^2_a}\right)^\frac{3}{2} e^{- \frac{E}{T}}.
\end{equation} 

The energy loss rates $Q_{a}$ via these two processes
are given by 
\begin{equation}
    Q_{a} =\int^{\infty}_{m_{a}} d\omega \,\omega \frac{d^{2} n_{a}}{dtd\omega} \Bigg|_{\rm{prim}} +  \int^{\infty}_{m_{a}} dE\,E \frac{d^{2} n_{a}}{dtdE} \Bigg|_{\rm{coal}} .
\label{eq:Qa}
\end{equation}
The photon coalescence contributes to the ALP production only when ALPs mass satisfies $m_a > 2\omega_{\rm{pl}}$.

We apply Eq. (\ref{eq:Qa}) to the models shown in Fig. \ref{fig:tmp-dns} and estimate the ALP cooling rate in post-processing.
Figure \ref{fig:Qa} shows the radial profile of ALP cooling rate $Q_a$ at $\tpb = 1\,\mathrm{s}$ for a coupling constant $\gten = g_{\alpha \gamma} / (10^{-10} \,\rm{GeV^{-1}}) = 10$, with the ALP mass $\ma=10\,\mathrm{MeV}$ (left panel) and $\ma=200\,\mathrm{MeV}$ (right panel). 
The cooling rates show peaks at $r \approx 7-10\,\mathrm{km}$, reflecting the strong temperature dependence of the ALP production rate and the fact that all rotating and non-rotating models have a temperature maximum at this radius. 
Figure \ref{fig:Qa} also shows that the ALP production is significantly suppressed by rotation. This is because the temperature is reduced by rotation, as shown in Fig. \ref{fig:tmp-dns}.

\section{Energy-loss argument}
\label{energy-loss argument}
ALPs produced inside a supernova can transport energy out of the PNS and serve as a cooling channel. Excessive ALP cooling would shorten the duration of the neutrino burst \cite{burrows_axions_1989,keil_fresh_1997}, making the neutrino signal predicted by SN simulations inconsistent with the SN~1987A observations by the Kamiokande-II \cite{hirata_observation_1987}, IMB experiments \cite{bionta_observation_1987}, and Baksan \cite{alekseev_possible_1987}. This idea forms the basis of the so-called ``energy-loss argument", which provides constraints on the ALP parameter space ($g_{a\gamma}-\ma$ plane). The most robust approach based on this argument would be to perform long-term, realistic SN simulations incorporating ALP energy transport for a variety set of ALP parameters and directly evaluate whether the duration is significantly shortened.
However, this method requires extensive parameter surveys and is too computationally expensive to be practical.

Therefore, in this study, we adopt a simplified approach based on prescription of Ref. \cite{raffelt_astrophysical_2008} to obtain constraints from relatively short-term simulations. 
In the standard core-collapse scenario, the PNS cooling is dominated by neutrino emission, and at $\tpb=1\,\s$ the neutrino emission has reached a quasi-steady cooling phase. Thus, to impose constraints on the ALP parameters, we estimate the ALP luminosity $L_a^{\mathrm{PNS}}$ through post-processing from the simulation data and evaluate whether it satisfies the following relation.
\begin{eqnarray}
    L_{a}^{\mathrm{PNS}} < L_{\nu}~~(\mathrm{at}~~\tpb=1\,\mathrm{s}).
\label{ELA_condition}
\end{eqnarray}
Here, $L_\nu$ is the total neutrino luminosity and $L_a^{\mathrm{PNS}}$ is defined by integrating the local ALP emissivity $Q_a$ over the entire PNS as
\begin{eqnarray}
L_{a}^{\mathrm{PNS}}=4\pi\int_0^{r_\mathrm{PNS}} Q_{a}r^2\,dr,
\label{L_a}
\end{eqnarray}
where $r_\mathrm{PNS}$ denotes the PNS radius.
If this condition is violated, ALP-induced energy loss effectively contributes to PNS cooling and significantly shortens the neutrino-burst duration compared to observations.

\begin{figure}[t]
    \centering
    \includegraphics[keepaspectratio, width=8.5cm,clip]{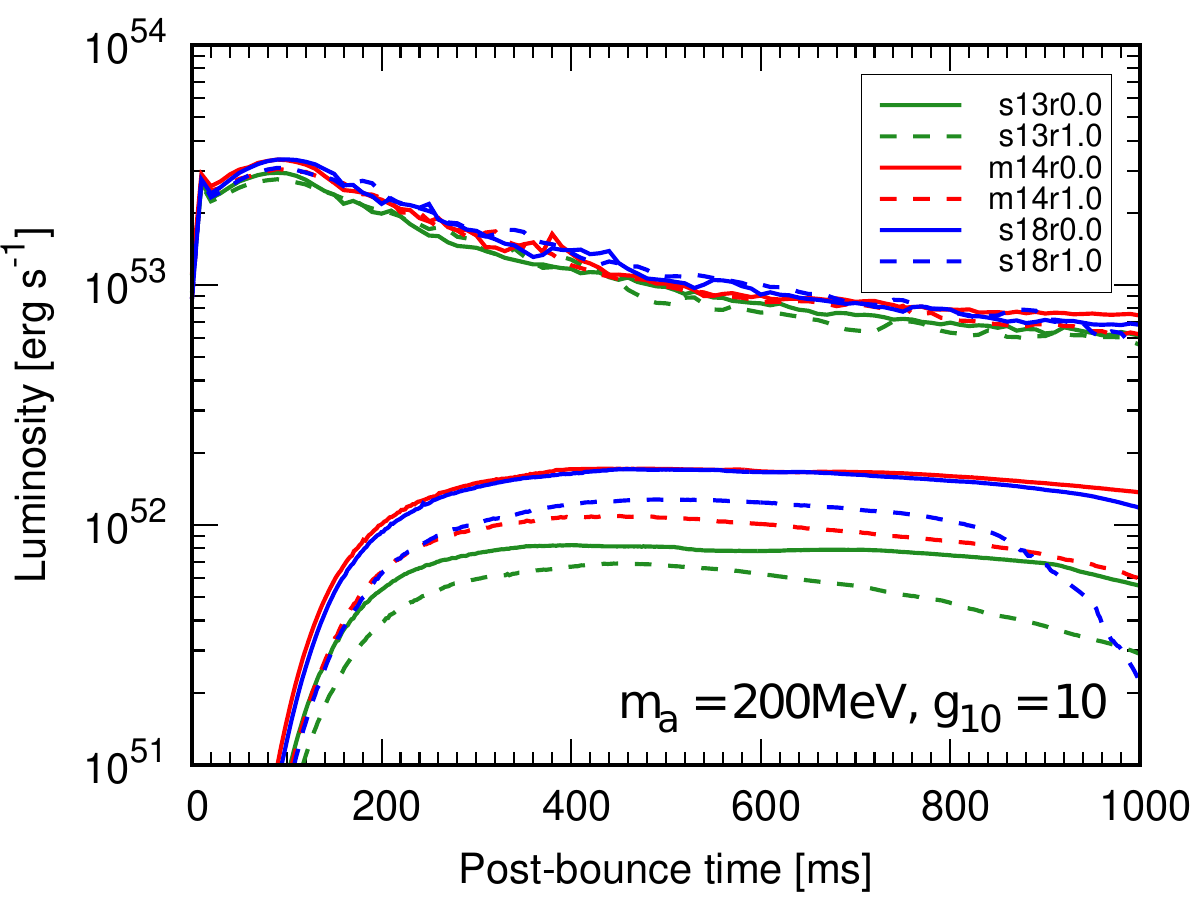}
    \caption{The time evolution of the neutrino luminosity and ALP cooling rate \(L_a\). At \(\tpb=1\) s, in all models, the neutrino luminosity is $6-8\times10^{52} \mathrm{\,erg \,s^{-1}}$, with only minor differences between progenitor models or the rotational parameters. The more massive progenitor models (m14; red lines and s18; blue lines) show a higher ALP cooling rate, but in rotating models, it is reduced compared with non-rotating ones. Notably, this reduction is most pronounced in the s18r1.0 model (blue dashed line).
    }
\label{fig:lum}
\end{figure}

The neutrino luminosity and the ALP luminosity $L_a^{\mathrm{PNS}}$ are affected by the progenitor models and rotation, as they lead to different PNS structures. To see these effects, we plot the neutrino and ALP luminosities for $g_{10}=10$ and $\ma = 200\,\MeV$ in Figure \ref{fig:lum}. 
The neutrino luminosities peak at $\tpb\sim 100\,\mathrm{ms}$, and then gradually decrease during the accretion phase. At $\tpb=1\,\mathrm{s}$, they reach $\sim 6-8 \times 10^{52}\,\mathrm{erg\,s^{-1}}$, with relatively minor dependence on the progenitor models or the rotation parameters. On the other hand, the ALP cooling rates are suppressed in the rotating models (dashed lines) compared to the non-rotating ones (solid lines). In particular, at $\tpb=1\,\s$, the cooling rate in the s18 model is reduced by approximately $80\,\%$. Since the ALP cooling rate is highly sensitive to the core temperature, this suppression follows the trend of the maximum temperature shown in Fig. \ref{fig:tmp-dns}. 

\begin{figure*}[t]
    \centering
    \includegraphics[keepaspectratio, width=17cm,clip]{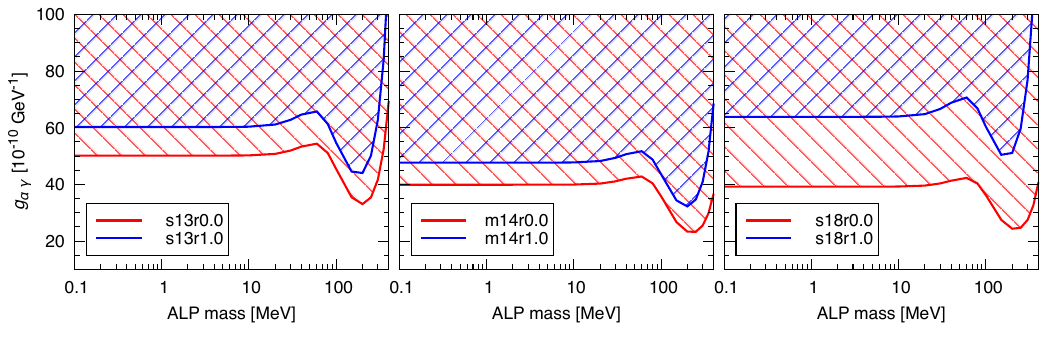}
    \caption{The ALP-excluded region in the \(g_{a\gamma}-m_a\) plane obtained from energy-loss argument. The red (blue) hatched region represent the parameter space excluded by the non-rotating (rotating) models. Rotation relaxes the constraints due to the reduction of the ALP cooling rate. In the s18 model, the rotating model shows a significant decrease in ALP cooling rate at \(\tpb=1\) s, making the impact of rotation on the excluded region particularly pronounced.
}
\label{fig:ELoss}
\end{figure*}

By adopting Eq. (\ref{ELA_condition}) as a criterion, we impose constraints on the MeV-scale ALP parameter space. Figure \ref{fig:ELoss} shows the excluded parameter space in the \(g_{a\gamma}-m_a\) plane for the six models. 
We find that rotation relaxes the constraints, as it lowers the PNS temperature and thereby reduces the ALP production rate. Among the three progenitors, the impact of rotation is most significant in the s18 model, while the effect is relatively modest in the s13 and m14 models. This difference arises from the stronger rotational suppression of the temperature at $\tpb=1\,\s$ in s18 model.
The constraints also reflect that the dominant ALP production process depends on the ALP mass. For $m_a \lesssim 80\,\mathrm{MeV}$, the Primakoff process dominates, whereas for $\ma \sim 80-200\,\mathrm{MeV}$, the photon coalescence becomes dominant channel. At $m_a \gtrsim 200\,\mathrm{MeV}$, the ALP production rate is strongly suppressed by the Boltzmann factor. This behavior has also been discussed in previous studies \cite[e.g.][]{lucente_heavy_2020}.

\section{Gamma-ray limit}
\label{gamma-ray limit}
When the coupling constant is sufficiently small ($g_{a\gamma}\lesssim10^{-10}\,\mathrm{GeV^{-1}}$ for MeV-scale ALP), ALPs produced in the SN core can freely escape from the SN without being impeded. They subsequently decay into two photons in the interstellar space. The daughter photons can have energies of $\gtrsim 10\,\mathrm{MeV}$, significantly exceeding the typical energy of photons from the stellar surface. 
For SN~1987A, gamma-ray observations were carried out by the gamma-ray spectrometer on board the Solar Maximum Mission (SMM) satellite, which was in operation at the time of the explosion. However, the detector did not detect any statistically significant excess of gamma rays \cite{chupp_experimental_1989}. 
Therefore, this non-detection places an upper limit on the gamma-ray fluence from ALP decays,
\begin{eqnarray}\label{eq:fcrit}
\mathcal{F}_\gamma< 1.78\,\gamma\cdot \mathrm{cm}^{-2} \, .
\end{eqnarray}
Comparing the gamma-ray fluence from ALP decays predicted by SN simulations with this observational upper limit allows us to constrain the ALP parameter space.

To estimate the gamma-ray fluence from ALPs produced in our SN simulations, we follow the method of Ref. \cite{jaeckel_decay_2018}; here we briefly outline the procedure.
The total ALP emission spectrum is obtained by integrating the production rate over the duration of the core collapse simulations \cite{payez_revisiting_2015},
\begin{align}
     \frac{dN_a}{dE_a}
&   = \int_{0}^{1\,\mathrm{s}} dt \,\frac{d\dot{N}_a}{dE_a} \notag \\
&   = \int^{1\,\s}_{0}dt\int^{r_\mathrm{SN}}_0 4\pi r^2dr\left[\omega\frac{d^{2} n_{a}}{dtd\omega} \Bigg|_{\rm{prim}} +  
    E\frac{d^{2} n_{a}}{dtdE} \Bigg|_{\rm{coal}} \right].
    \label{eq:spectrum}
\end{align}
Here, $r_{\mathrm{SN}}=20,000\,\km$ denotes the spatial domain of our SN simulations, and the time integration is performed up to $\tpb = 1\,\mathrm{s}$, corresponding to the duration of our simulations.

\begin{figure*}[t]
    \centering
    \includegraphics[keepaspectratio, width=17cm,clip]{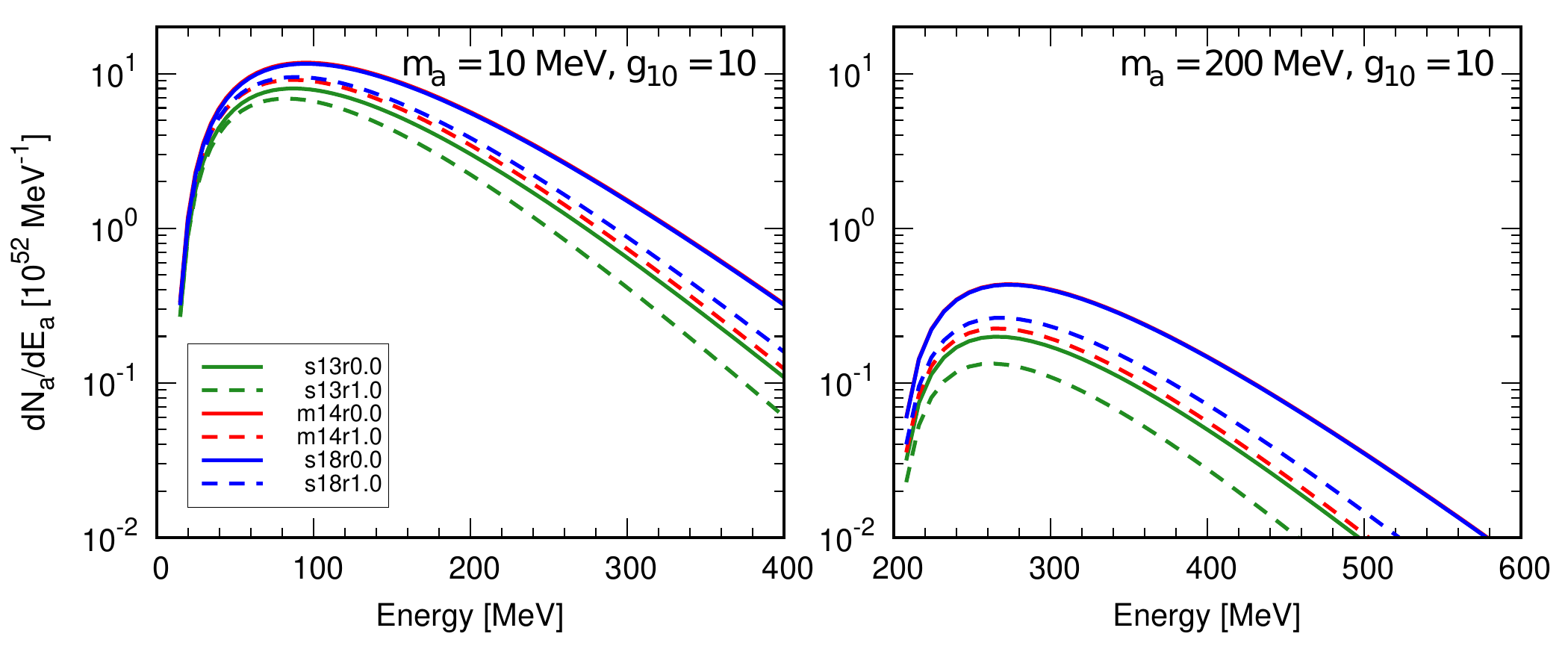}
    \caption{ALP energy distribution from Eq.\ref{eq:spectrum}. The color coding is the same as in Fig. (\ref{fig:Qa}). The total emission spectrum is suppressed by rotation for all progenitor models, with the suppression becoming stronger at high energies and being most pronounced for the m14 model, which exhibits the largest temperature decrease throughout $\tpb = 1\,\mathrm{s}$.
    The m14r0.0 and s18r1.0 models show nearly identical ALP energy distributions and therefore overlap.
    }
\label{fig:spectrum}
\end{figure*}

Figure \ref{fig:spectrum} shows the ALP energy distribution from Eq. (\ref{eq:spectrum}).
As representative cases, ALPs with two different masses, $m_a = 10~\mathrm{MeV}$ (left panel) and $m_a = 200~\mathrm{MeV}$ (right panel), are shown with the coupling constant fixed at $g_{10}=10$. 
For an ALP, its energy $E_a$ obeys the relativistic kinematic relation $E_a^2 = |\mathbf{p}_a|^2 + m_a^2$ and therefore must satisfy $E_a \ge m_a$. Consequently, the energy spectrum exhibits a sharp cutoff at $E_a = m_a$.
For all progenitor models, the total emission spectrum is suppressed by rotation, particularly at high-energy regime. This rotational suppression is most pronounced in the m14 progenitor model, because in this model the temperature remains lower due to rotation for a relatively long period, and the total energy spectrum reflects the time-integrated ALP emission.

To constrain the ALP parameter space with the non-detection of gamma-rays from SN~1987A, we estimate the gamma-ray fluence expected from our SN simulations. Here, we focus on the m14 progenitor model, which can reproduce observational features of the SN~1987A progenitor.
Assuming that all ALPs produced inside the SN decay into photons outside the SN and in its vicinity, the observed fluence is given by
\begin{eqnarray}\label{eq:fgame}
    \mathcal{F}_{\gamma}|_{\mathrm{Earth}} = \frac{2}{4\pi d^{2}_{\mathrm{SN}}}\int \frac{d{N}_a}{dE_{a}}{dE_{a}},
\end{eqnarray}
where $d_\mathrm{SN}=51.4 \,\mathrm{kpc}$ is the distance between the Earth and SN~1987A.
However, to estimate the fluence $\mathcal{F}^\mathrm{exp}_{\gamma}$ actually expected to be observed at Earth, it is necessary to adopt the following expression after taking several effects into account.
\begin{eqnarray}
\mathcal{F}^\mathrm{exp}_{\gamma}=\mathcal{F}_{\gamma}|_{\mathrm{Earth}} \times\mathcal{P_{\mathrm{survive}}}\mathcal{P_{\mathrm{decay}}}\mathcal{P_{\mathrm{time}}}.
\label{eq:Fexp}
\end{eqnarray}

$P_{\mathrm{survive}}$ is the probability that an ALP escapes from the progenitor without decaying inside the SN, 
\begin{eqnarray}
    \mathcal{P_{\mathrm{survive}}} = \exp\left[-\frac{R_\mathrm{star}}{l_\mathrm{ALP}(E_a)}\right],
\end{eqnarray}
where $R_{\mathrm{star}} \sim 3\times10^{10}\,\mathrm{m}$ is the effective stellar radius \cite{Kazanas_supernova_2015}.
$l_{\mathrm{ALP}}(E_a)$ is the ALP decay length, given by
\begin{align}
l_\mathrm{ALP} &= \frac{\gamma v}{\Gamma_{a\gamma\gamma}} 
= \frac{E_a}{m_a}\sqrt{1-\frac{m_a^2}{E_a^2}}\frac{64\pi}{g^2_{a\gamma\gamma}m_a^3} \notag \\
&\approx 4\times10^{13}\,\mathrm{m} \notag \\
&\times\left(\frac{E_a}{100\,\mathrm{MeV}}\right)
\left(\frac{10\,\mathrm{MeV}}{m_a}\right)^4\left(\frac{10^{-10}\,\mathrm{GeV^{-1}}}{g_{a\gamma\gamma}}\right)^2.
\end{align}

$P_{\mathrm{decay}}$ is a factor that accounts for the fact that, when an ALP decays beyond the distance between the Earth and SN~1987A, the daughter photons are typically not detected.
$P_{\mathrm{time}}$ denotes the fraction of daughter photons that arrive within the detector’s measurement time window.
These two factors are somewhat entangled, and the probability of satisfying these conditions is given by
\begin{eqnarray}\label{eq:pdpt}
    \mathcal{P_{\mathrm{decay}}} \times \mathcal{P_{\mathrm{time}}}\approx \delta tE_ag^2_{a\gamma\gamma}m^2_a
\end{eqnarray}
where $\delta t\simeq223\,\s$ is the full time window of the SMM gamma-ray spectrometer \cite{chupp_experimental_1989}.

In addition to these factors, for SN~1987A we consider the gamma-ray energy range
$E_\gamma = [25,\,100]\,\mathrm{MeV}$.
Since $E_\gamma \simeq E_a/2$, the detector sensitivity therefore corresponds to ALPs
in the energy range $E_a = [50,\,200]\,\mathrm{MeV}$.
Therefore, taking these effects into account, the fluence of ALP-originated photons must satisfy the upper limit obtained by the SMM observation of SN~1987A:
\begin{eqnarray}\label{eq:fcrit}
\mathcal{F}_\gamma^{\mathrm{exp}}(223\,\mathrm{s})
< 1.78\,\gamma\cdot \mathrm{cm}^{-2} \, .
\end{eqnarray}

\begin{figure}[t]
    \centering
    \includegraphics[keepaspectratio, width=8.5cm,clip]{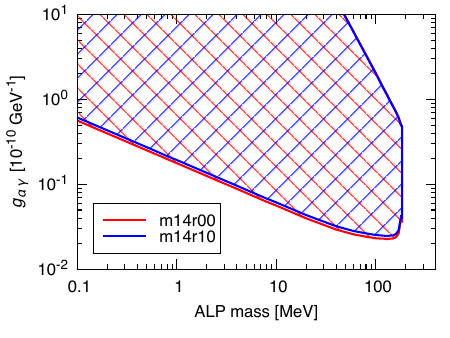}
    \caption{The ALP-excluded region in the $g_{a\gamma}-\ma$ plane obtained from gamma-ray limit for the m14 model. The color coding is the same as in Fig. \ref{fig:ELoss}. For the m14 model, the constraint is slightly relaxed by rotational effects, while for the other two progenitor models the relaxation of the constraints is negligible.}
\label{fig:gray}
\end{figure}

We apply equations (\ref{eq:fgame})--(\ref{eq:pdpt}) to our m14 models to estimate the expected gamma-ray fluence and employ equation (\ref{eq:fcrit}) to derive constraints on the ALP parameters. Figure \ref{fig:gray} shows the constraints derived from the gamma-ray limit for the m14 progenitor model corresponding to SN~1987A.
One can find that the relaxation of the gamma-ray limit due to rotation is modest for the m14 model. This is because the expected gamma-ray fluence scales as $\mathcal{F}_\gamma^{\mathrm{exp}}\propto g_{10}^4$ because of $dN_a/dE_a \propto g_{10}^2$ and equation (\ref{eq:pdpt}). This relation implies that, even if the gamma-ray fluence is reduced by half due to rotation, the resulting constraint on $g_{10}$ changes only by $\sim 16\%$. We also confirm that the effect of rotation on the gamma-ray limits is negligible for the s13 and s18 models.

\section{Conclusion and Discussion}
\label{CandD}

Rotation in CCSNe has been extensively studied, yet its impact on ALP emission and  the resulting constraints has been poorly investigated.
An existing study of ALP emission from rotating SNe \cite{bar_there_2020} is based on very rapid progenitor rotation and a collapse-induced thermonuclear explosion mechanism. A systematic investigation of ALP emission in typical CCSN models is still lacking.
In this study, we perform two-dimensional core-collapse supernova simulations with rotation to quantify the impacts of these effects and their dependence on progenitor properties.
We employed $13.0\,\Msun$ and $18.0\,\Msun$ single-star models and a $14+9\,\Msun$ merger model based on the slow-merger scenario as representative progenitors with different masses and internal structures. In addition, we considered rotation with the angular velocity of $\Omega_0=0.0$ and $1.0\,\mathrm{rad\,s^{-1}}$. From these simulations, we evaluate the ALP emission rates and derive constraints on MeV-scale ALP parameters based on the energy-loss argument \cite{raffelt_astrophysical_2008} and the gamma-ray limit \cite{jaeckel_decay_2018}. 

The structural differences induced by the progenitor model and rotation play a crucial role in determining the central temperature of the SN core. In all models, the temperature peaks at $r \simeq 7-10\,\mathrm{km}$ within $\tpb \le1\,\s$. 
The the maximum temperature is reduced for the rotating models since the centrifugal support decreases the gravitational energy released during mass accretion. The magnitude of the temperature reduction due to rotation depends on the progenitor models and the reduction is particularly pronounced in the high-compactness m14 and s18 models. Notably, the rotating s18 model exhibits a remarkable decrease in temperature at $\tpb \simeq 0.9-1\,\mathrm{s}$. Since ALP emission rates are highly sensitive to temperature, this temperature reduction due to rotation leads to a substantial suppression of ALP emission in the m14 and s18 models.

We find that constraints derived from the energy-loss argument, $L_a < L_\nu$ at $\tpb = 1\,\s$, are relaxed by rotation through the suppression of ALP cooling, and that this relaxation is most significant in high-compactness s18 models. In contrast, for the SN~1987A gamma-ray limit, the constraint relaxation is negligible, despite rotation suppressing the total ALP emission.

A representative neutrino luminosity ($L_\nu \simeq 2-3\times10^{52}\,\mathrm{erg\,s^{-1}}$ at $\tpb=1\,\s$) has commonly been adopted in energy-loss argument studies \cite[e.g.,][]{Chang_supernova_2018, bollig_muons_2020, carenza_constraints_2020, lucente_supernova_2021, lucente_constraining_2022, Ferreira_strong_2022}.
In this work, we instead use the neutrino luminosity obtained from our numerical simulations ($L_\nu \sim 6-8\times10^{52}\,\mathrm{erg\,s^{-1}}$).
The neutrino luminosities predicted by our CCSN models are higher than those adopted in previous studies, because we employ a relatively soft EOS (Lattimer and Swesty EOS with $K=220$\,MeV \cite{lattimer_generalized_1991}) as well as different progenitor models. 
The PNS contraction and mass-accretion history, which strongly affect neutrino emission properties, depend on the EOS and the progenitor structure. 
Hence, the criterion adopted in this work, which compares the ALP emission with the total neutrino luminosity,  depends on the internal structure and the initial angular velocity of the underlying SN model.

In addition to rotational effects, the ALP constraints can be affected by several other factors.
In this work, we neglected ALP feedback effects due to the need to survey a wide range of ALP parameters, which entails high numerical cost.
In our previous works~\cite{mori_shock_2022, mori_multimessenger_2023, Takata_progenitor_2025}, we performed SN simulations incorporating ALP transport and reported that ALPs heat the matter behind the shock, thereby promoting shock expansion.
The enhanced shock expansion reduces the gravitational energy released during mass accretion, which in turn lowers the PNS temperature.
Since the ALP production rate is highly sensitive to temperature, this leads to a suppression of the ALP emissivity. Consequently, for the energy-loss argument, ALP feedback effects are expected to relax the constraints. In contrast, for the gamma-ray limit, the coupling constants are too weak to affect SN dynamics, and the constraints are therefore expected to remain largely unchanged.

Another source of uncertainty is the numerical resolution in simulations.  Ref.~\cite{nagakura_towards_2019} indicated that models with higher resolution can resolve turbulence more accurately, which enhances the neutrino heating rate. This promotes shock expansion and, through the same mechanism described above, is expected to further relax the ALP constraints.

The simple prescription of the energy-loss argument ($L_a<L_\nu$ at $\tpb=1.0\,\s$) has been widely adopted in previous studies. Using this prescription, the s18r1.0 model exhibits a significantly relaxed constraint due to rotation, since the temperature drops at $\tpb\simeq0.9-1.0\,\s$, leading to a substantial reduction in the ALP cooling rate. However, the timing of such a temperature decrease can depend on the model setup. This suggests that fixing the evaluation time at $\tpb=1\,\s$ can introduce arbitrariness in the constraints.
A more robust approach is to directly assess, in long-term SN simulations which incorporate ALP energy transport, the extent to which the neutrino duration is shortened by ALP cooling. Three-dimensional (3D) simulations are also needed to draw a robust conclusion to rotational effects including non-axisymmetric instabilities that deal with magnetic fields more accurately \cite{kuroda20,martin2021,shibagaki_21,matteo23,shibagaki_2026}. This motivates the need to make possible extended CCSN simulations in 3D beyond $\tpb > 1\,\s$ coupled with ALP transport.

\begin{acknowledgments} 
Numerical computations were carried out on the Cray XD2000 at the Center for Computational Astrophysics, National Astronomical Observatory of Japan. This work is supported by JSPS KAKENHI Grant Numbers JP23KJ2147, JP23K13107, JP23K20862, JP23K22494, JP24K00631, JP25H02194, JP26K07093, and funding from Fukuoka University (Grant No.GR2606) and also by MEXT as “Program for Promoting researches on the Supercomputer Fugaku” (Structure and Evolution of the Universe Unraveled by Fusion of Simulation and AI; Grant Number JPMXP1020230406) and JICFuS.
\end{acknowledgments}
\bibliographystyle{apsrev4-2}
\bibliography{takata}
\end{document}